\documentclass[apj,numberedappendix,revtex4,twocolappendix]{emulateapj}
\usepackage{natbib}
\usepackage{graphicx}
\usepackage[toc,page]{appendix}
\usepackage{natbib}
\usepackage{color}
\usepackage{threeparttable}
\usepackage{hyperref}
\usepackage{breakurl}
\usepackage{float}
\usepackage[caption = false]{subfig}
\usepackage[normalem]{ulem}

\begin{document}

%\slugcomment{Draft version: \today}
\slugcomment{Accepted for publication in the Astrophysical Journal Letters}
\title{Galaxy Structure as a Driver of the Star Formation Sequence Slope and Scatter} 
\email{kate.whitaker@nasa.gov}
\author{Katherine E. Whitaker\altaffilmark{1,10}, Marijn Franx\altaffilmark{2}, 
Rachel Bezanson\altaffilmark{3,11}, Gabriel B. Brammer\altaffilmark{4}, \\
Pieter G. van Dokkum\altaffilmark{5}, Mariska T. Kriek\altaffilmark{6}, Ivo Labb\'{e}\altaffilmark{2}, Joel Leja\altaffilmark{5}, 
Ivelina G. Momcheva\altaffilmark{5}, Erica J. Nelson\altaffilmark{5}, \\  Jane R. Rigby\altaffilmark{1}, Hans-Walter Rix\altaffilmark{7},
Rosalind E. Skelton\altaffilmark{8}, Arjen van der Wel\altaffilmark{7}, Stijn Wuyts\altaffilmark{9}} 
\altaffiltext{1}{Astrophysics Science Division, Goddard Space Flight Center, Code 665, Greenbelt, MD 20771, USA}
\altaffiltext{2}{Leiden Observatory, Leiden University, P.O. Box 9513, 2300 RA Leiden, The Netherlands}
\altaffiltext{3}{Steward Observatory, Department of Astronomy, University of Arizona, AZ 85721, USA}
\altaffiltext{4}{Space Telescope Science Institute, 3700 San Martin Drive, Baltimore, MD 21218, USA}
\altaffiltext{5}{Department of Astronomy, Yale University, New Haven, CT 06520, USA}
\altaffiltext{6}{Astronomy Department, University of California, Berkeley, CA 94720, USA}
\altaffiltext{7}{Max Planck Institut fur Astronomie, K\"{o}nigstuhl 17, D-69117 Heidelberg, Germany}
\altaffiltext{8}{South African Astronomical Observatory, PO Box 9, Observatory, Cape Town, 7935, South Africa}
\altaffiltext{9}{Max-Planck-Institut für extraterrestrische Physik, Postfach 1312, Giessenbachstr, D-85741 Garching, Germany}
\altaffiltext{10}{NASA Postdoctoral Program Fellow}
\altaffiltext{11}{Hubble Fellow}
\shortauthors{Whitaker et al.}
%\shorttitle{}

\begin{abstract}
It is well established that (1) star-forming galaxies follow a relation between their star 
formation rate (SFR) and stellar mass (M$_{\star}$), the ``star-formation sequence'', and (2)  
the SFRs of galaxies correlate with their structure, where star-forming galaxies are less 
concentrated than quiescent galaxies at fixed mass. Here, we consider whether the scatter and slope of the 
star-formation sequence is correlated with systematic variations in the S\'{e}rsic indices, $n$, of 
galaxies across the SFR-M$_{\star}$ plane.
We use a mass-complete sample of 23,848 galaxies at 0.5$<$$z$$<$2.5 selected from the 3D-HST
photometric catalogs.  Galaxy light profiles parameterized by $n$ are based on 
\emph{Hubble Space Telescope} CANDELS near-infrared imaging.  We use
a single SFR indicator empirically-calibrated from stacks of \emph{Spitzer}/MIPS 24$\mu$m imaging, adding the unobscured and 
obscured star formation.
We find that the scatter of the star-formation sequence is related in part to galaxy structure; the scatter due to variations 
in $n$ at fixed mass for star-forming galaxies ranges from 0.14$\pm$0.02 dex at $z$$\sim$2 to 0.30$\pm$0.04 dex at $z$$<$1.
While the slope of the $\log\mathrm{SFR}-\log\mathrm{M_{\star}}$ relation is of order unity for disk-like galaxies,
galaxies with $n$$>$2 (implying more dominant bulges) have significantly lower $\mathrm{SFR/M_{\star}}$ than the main ridgeline of the
star-formation sequence. 
These results suggest that bulges in massive $z$$\sim$2 galaxies are actively building up,
where the stars in the central concentration are relatively young.
At $z$$<$1, the presence of older bulges within star-forming galaxies lowers global $\mathrm{SFR/M_{\star}}$, decreasing the 
slope and contributing significantly to the scatter of the star-formation sequence. 
\end{abstract}


\section{Introduction}
\label{sec:intro}

One of the outstanding problems in galaxy formation is understanding the 
causal relationship between the morphologies and star-formation histories of galaxies.
There is increasing evidence that galaxies with quiescent stellar populations
have significantly smaller sizes and more concentrated light profiles
than actively star-forming galaxies at fixed stellar mass out to $z=2.5$
 \citep{Kriek09b,Williams10,Wuyts11b,vanderWel14}.  
We know that quenching occurs, causing one population to migrate to the other.  However,
the process(es) that are primarily responsible for the shutdown of star formation are not well 
understood.  It is yet unclear if the well-defined correlation between stellar mass and star formation 
rate \citep[SFR; e.g.,][and numerous others]{Brinchmann04, Noeske07a, Whitaker14b, Schreiber15}
 is driven by galaxy mass alone, or if some other parameter, such as surface density \citep{Franx08}
 or bulge mass \citep{Lang14}, comes into play.  

In the nearby universe, \citet{Abramson14} find that increasing the galaxy bulge mass-fraction lowers
the global specific SFR ($\mathrm{sSFR}\equiv\mathrm{SFR/M_{\star}}$).  
Their working assumption is that star formation occurs in disks \citep{Kennicutt98},
and bulges are composed primarily of older stars.  While the bulge contributes to the total stellar mass of the galaxy,
it does not significantly contribute to the global star formation, thereby depressing the sSFR.
This naturally predicts a flattening of the galaxy $\log\mathrm{SFR}-\log\mathrm{M}_{\star}$
relation at the massive end where galaxies are bulge-dominated.   
\citet{Whitaker14b} measure a strong evolution in the slope of the star formation sequence at the massive end, 
consistent with a flat slope at $z$$<$0.5 and a slope close to unity at $z$$>$2.5. In other words, even
at $z$$\sim$2 we are seeing a depression in the global specific SFRs amongst massive galaxies. 

\citet{Lang14} decomposed the bulges and disks of massive high-redshift galaxies out to $z$=2.5.
They speculate that significant bulge
growth precedes the departure from the star formation sequence. 
Lang et al. further demonstrate that the 
bulge mass is a more reliable predictor of quiescence than the total stellar mass \citep[see also][]{Bell12}.
To understand the regulation and quenching of star formation in 
galaxies across cosmic time, we must take these analyses one step further and connect the 
structural evolution of quiescent and star-forming galaxies with their sSFRs.

Before we can understand the details of how galaxies quench their star-formation, we must first 
synthesize the properties of massive galaxies in a self-consistent manner.  
Such an analysis requires a robust quantification of the rest-frame optical galaxy structural properies and 
a single indicator for the total obscured and unobscured star formation in galaxies for the entire population.  
In this letter, we combine S\'{e}rsic index measurements derived from CANDELS/WFC3
imaging with total UV+IR SFRs, using Spitzer/MIPS 24$\mu$m photometry, extending
the 24$\mu$m stacking technique presented in \citet{Whitaker14b} to probe a range in structural
parameters.  
We present the trends between galaxy M$_{\star}$, total sSFR, and
$n$ at 0.5$<$$z$$<$2.5, exploring the dependence of the slope and scatter
of the star-formation sequence on galaxy concentration.  

In this letter, we use a \citet{Chabrier} initial mass function and assume a $\Lambda$CDM 
cosmology with $\Omega_{\mathrm{M}}=0.3$, 
$\Omega_{\Lambda}=0.7$, and $\mathrm{H_{0}}=70$ km s$^{-1}$ Mpc$^{-1}$. All magnitudes are given in the AB system.

\section{Data and Sample Selection}
\label{sec:data}

\subsection{Stellar Masses, Redshifts and Rest-frame Colors}

We take advantage of the HST/WFC3 and ACS photometric and spectroscopic datasets in five 
well-studied extragalactic 
fields through the Cosmic Assembly Near-IR Deep Extragalactic Legacy Survey \citep[CANDELS;][]{Grogin11, Koekemoer11} and the
3D-HST survey \citep{Brammer12}.  
Using stellar masses, redshifts, and rest-frame colors from the 3D-HST 0.3--8$\mu$m photometric 
catalogs \citep[see][]{Skelton14}, we select samples of 9400, 8278, and 
6170 galaxies greater than stellar masses 
of $\log\mathrm{M_{\star}/M_{\odot}}=8.8$, 9.0, and 9.6
in three redshift intervals of 0.5$<$$z$$<$1.0, 1.0$<$$z$$<$1.5, and 1.5$<$$z$$<$2.5. 
The galaxies are split into star-forming and quiescent sub-samples based on their rest-frame 
$U$--$V$ and $V$--$J$ colors, following a modified definition of \citet{Whitaker12a}\footnote{We no longer
require $V$--$J$$<$1.5, similar to \citet{vanderWel14}, as this is a false upper limit imposed on the quiescent 
population. This effects $<1\%$ of galaxies.}.
We have used the Spitzer/IRAC color selections presented in
\citet{Donley12} to identify and remove luminous AGNs; 
3\% of the sample were removed as AGN candidates.  We note that the results of 
this letter do not depend on this step. The final sample comprises 23,848 galaxies at 0.5$<$$z$$<$2.5.

We indicate the stellar mass limits for which the majority of objects in the sample
have $\mathrm{H_{F160W}}<23.5$ (see Section~\ref{sec:morphology}).  These limits are well
above the mass-completeness limits presented in \citet{Tal14}, which are determined by comparing
object detection in CANDELS/deep with a re-combined subset of exposures reaching the CANDELS/wide depth.
The stellar masses of star-forming galaxies 
have been corrected for contamination of the broadband fluxes from emission lines using the values presented in Appendix A of \citet{Whitaker14b}.  
These corrections only become significant at $\log\mathrm{M_{\star}/M_{\odot}}<9.5$ and $z>1.5$.
For example, \citet{Whitaker14b} find that stellar masses of $\log\mathrm{M_{\star}/M_{\odot}}=9.5$ are 
overestimated by 0.06 (0.2) dex at $1.5<z<2.0$ ($2.0<z<2.5$) because of emission line fluxes.

Where available, we combine the spectral energy distributions (SEDs) with low-resolution HST/WFC3 G141 grism spectroscopy to derive
grism redshifts with $\Delta z/(1+z)=0.3\%$ accuracy \citep{Brammer12}. 
We select the ``best'' redshift to be the spectroscopic redshift, grism redshift or the photometric redshift, in this ranked order depending on availability.
Photometric redshifts comprise 62\% (79\%) of the $0.5<z<1.5$ ($1.5<z<2.5$) sample, while 26\% (19\%) have grism redshifts and 12\% (2\%) spectroscopic redshifts.

\subsection{Morphology}
\label{sec:morphology}

The \citet{Sersic68} index is a measure of the shape of the surface brightness profile of a galaxy,
indicating the concentration of
the light; it has been shown to correlate with quiescence \citep[e.g.,][]{deVaucouleur,Caon93}.
At 1$<$$z$$<$3, \citet{Bruce14} demonstrate that $n$=4 corresponds to a bulge-to-disk ratio
of about four in massive galaxies, where $\sim$80$\%$ of the light is in the bulge.  Exponential disks have $n$=1 and
bulge-to-disk ratios of about a quarter, with $20\%$ of the light in the bulge. The transition from bulge-dominated to
disk-dominated occurs around $n$=2.
We caution that we are not directly measuring the bulge mass-fractions, rather adopting $n$ as
a proxy for bulge dominance.
Furthermore, it is not clear whether this interpretation of $n$ holds at $z$$\gtrsim$1.

$n$ is measured from HST/WFC3 $J_{\mathrm{F125W}}$ 
at $0.5<z<1.5$ and $H_{\mathrm{F160W}}$ at $1.5<z<2.5$ \citep{vanderWel12}.
Simulations by \citet{vanderWel12} show that the systematic errors in
$n$ for all galaxy types are $<10\%$ for $\mathrm{H_{F160W}}<23.5$, thus motivating our mass limits.
It is complicated to correct for the wavelength dependence of $n$ within the redshift bins,
we do not correct for this effect but note that $n$ increases by $\sim$10\% from 4500$\mathrm{\AA}$ to
8000$\mathrm{\AA}$.

%=== Fig 1
\begin{figure*}[t!]
\leavevmode
\centering
\includegraphics[width=0.82\linewidth]{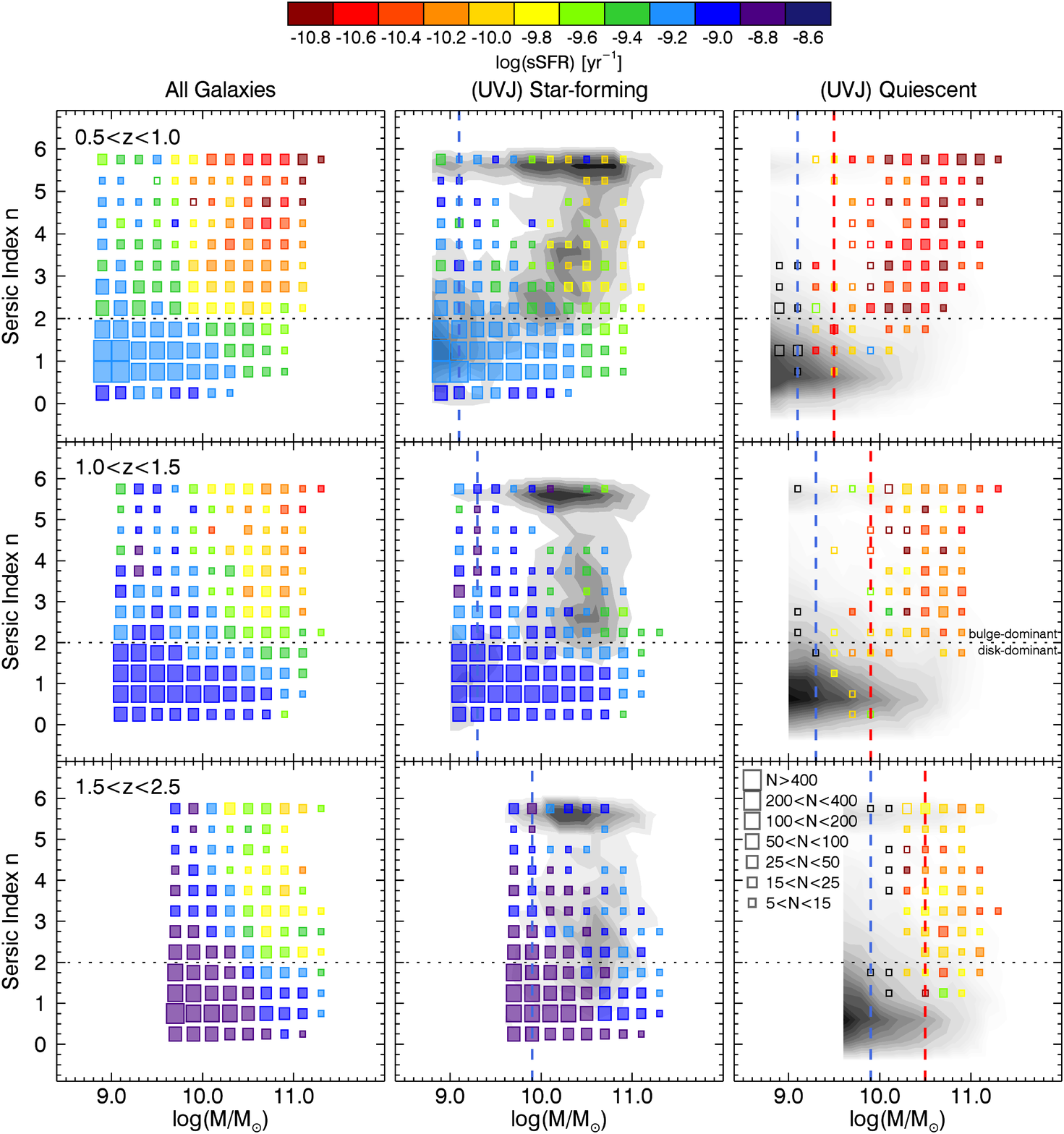} 
\caption{S\'{e}rsic indices of galaxies as a function of stellar mass, color-coded by the sSFRs derived
from UV+IR stacking analyses in 0.2 dex bins of $\log\mathrm{M_{\star}/M_{\odot}}$ and 0.5 width bins of $n$.
The vertical dashed lines correspond to the stellar mass limits down to which $n$ is robust for color-selected star-forming (blue) and
quiescent (red) populations.
At fixed stellar mass, galaxies with higher $n$ (implying prominent bulges) and $\log\mathrm{(M/M_{\odot})}>10$
have lower sSFRs, suggesting that the bulge formation is linked to quiescence.
The size of the symbol depends on the number of galaxies, and the underlying greyscale demarcates the opposite population (e.g.,
star-forming galaxies in the quiescent panels).}
\label{fig:sersic}
\end{figure*}

\subsection{Total Star Formation Rates}
Total star formation rates are derived from stacking analyses of Spitzer/MIPS 24$\mu$m photometry, following
the procedure detailed in \citet{Whitaker14b}\footnote{Stacking is currently the only way to derive SFRs from a single indicator 
at low stellar masses or SFRs, separating this work from \citet{Wuyts11b}.}.  
The Spitzer/MIPS 24$\mu$m images in the AEGIS field 
are provided by the Far-Infrared Deep Extragalactic Legacy (FIDEL) survey \citep{Dickinson07}, COSMOS from
the S-COSMOS survey \citep{Sanders07}, GOODS-N and GOODS-S from \citet{Dickinson03}, and 
UDS from the Spitzer UKIDSS Ultra Deep 
Survey\footnote{\url{http://irsa.ipac.caltech.edu/data/SPITZER/SpUDS/}}(SpUDS; PI: J. Dunlop).  
The analysis code uses a high-resolution $J_{\mathrm{F125W}}$+$H_{\mathrm{F140W}}$+$H_{\mathrm{F160W}}$ 
detection image as a prior to model the contributions from neighboring blended sources in the lower resolution 
MIPS 24$\mu$m image \citep{Labbe10}.  All galaxies are ``cleaned'' of the contaminating flux of the neighboring sources before
stacking. We refer the reader to Section 3 of \citet{Whitaker14b} for the full details of the MIPS 24$\mu$m stacking
analyses. The SFRs derived for quiescent galaxies herein are upper limits, as the 24$\mu$m technique
overestimates the SFRs for galaxies with $\log\mathrm{sSFR}$$<$$-10$ yr$^{-1}$, but remains robust above this limit \citep{Utomo14, Hayward14}.  For this reason, identifying star-forming galaxies from 24$\mu$m-based
sSFRs results in a significant fraction of interloping quiescent galaxies.

\section{Results}
\label{sec:sersic}

If increasing bulge mass-fractions result in a flattening of the star formation sequence at the massive end \citep{Abramson14},
we should see a correlation between the $n$, M$_{\star}$, and sSFR.
In Figure~\ref{fig:sersic}, we show the S\'{e}rsic indices of galaxies as a function of their stellar mass, color-coded by the
UV+IR sSFRs for all galaxies (left), star-forming only (middle), and quiescent only (right).
Indeed, we find a decrease in the sSFRs for galaxies with high $n$ at $\log\mathrm{M_{\star}/M_{\odot}}>10$.
There does not appear to be any dependence of $n$ on the average
sSFR for low-mass galaxies with $\log\mathrm{M_{\star}/M_{\odot}}<10$.

Quiescent galaxies with $\log\mathrm{(M/M_{\odot})}>10$ tend to have $n$$>$2,
consistent with bulge-dominated light profiles.  Although this finding is not new, we do show for the first
time that the analogous star-forming sub-population with similar $n$-values
also have depressed sSFRs relative to the bulk of the star-forming population. 
We also see a low-mass quiescent population at 0.5$<$$z$$<$1.0 with similar (disk-like) S\'{e}rsic indices 
in Figure~\ref{fig:sersic}, albeit these galaxies lie close to or below the mass limit.
  This may suggest that lower-mass galaxies quench their star formation at later times through
a more gradual ``slow-track'' of gas depletion \citep[see, e.g.,][]{Barro13}.

%=== Fig 2
\begin{figure}[t!]
\leavevmode
\centering
\includegraphics[width=\linewidth]{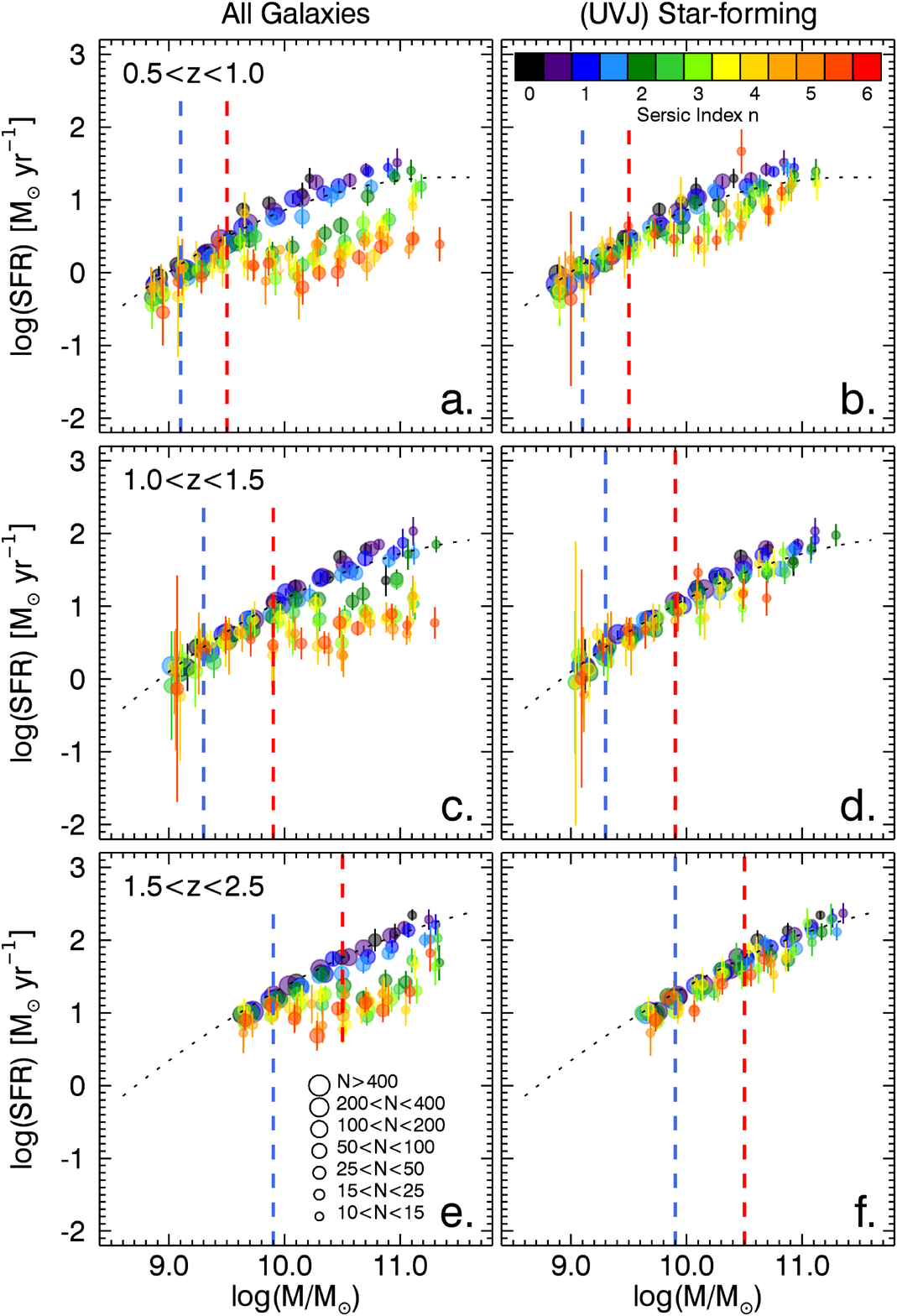}
\caption{The $\log(\mathrm{SFR})$--$\log\mathrm{M_{\star}}$ relation of all galaxies (left) and color-selected 
star-forming (right) resulting from the stacking analysis across
the $n$-M$_{\star}$ plane and vertical lines defined in Figure~\ref{fig:sersic}.  Each symbol is color-coded by $n$;
the size of the symbol depends on the number of galaxies in the bin.
The error bars are derived from a Monte Carlo bootstrap simulation of the stacking analyses. 
The average star formation sequence from \citet{Whitaker14b} is shown with dotted lines.
Galaxies with significant bulges ($n$$>$2) exhibit a shallower slope for the star formation-stellar
mass relation.}
\label{fig:sersic_sfr}
\end{figure}

%%=== Fig 3
\begin{figure*}[t!]
\subfloat[All Galaxies]{\includegraphics[width=.48\linewidth]{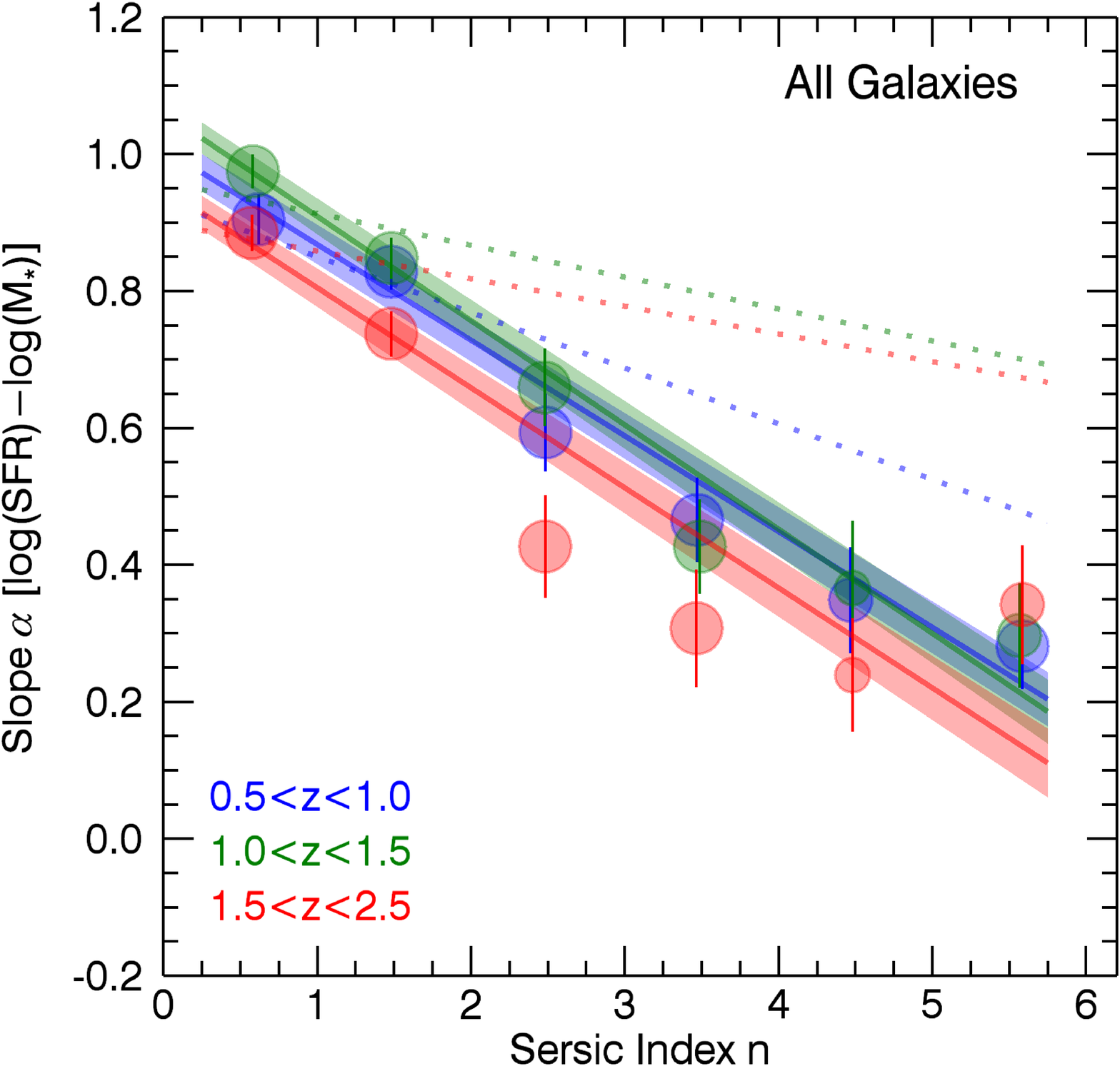}}\hfill
\subfloat[(UVJ) Star-forming Galaxies]{\includegraphics[width=.48\linewidth]{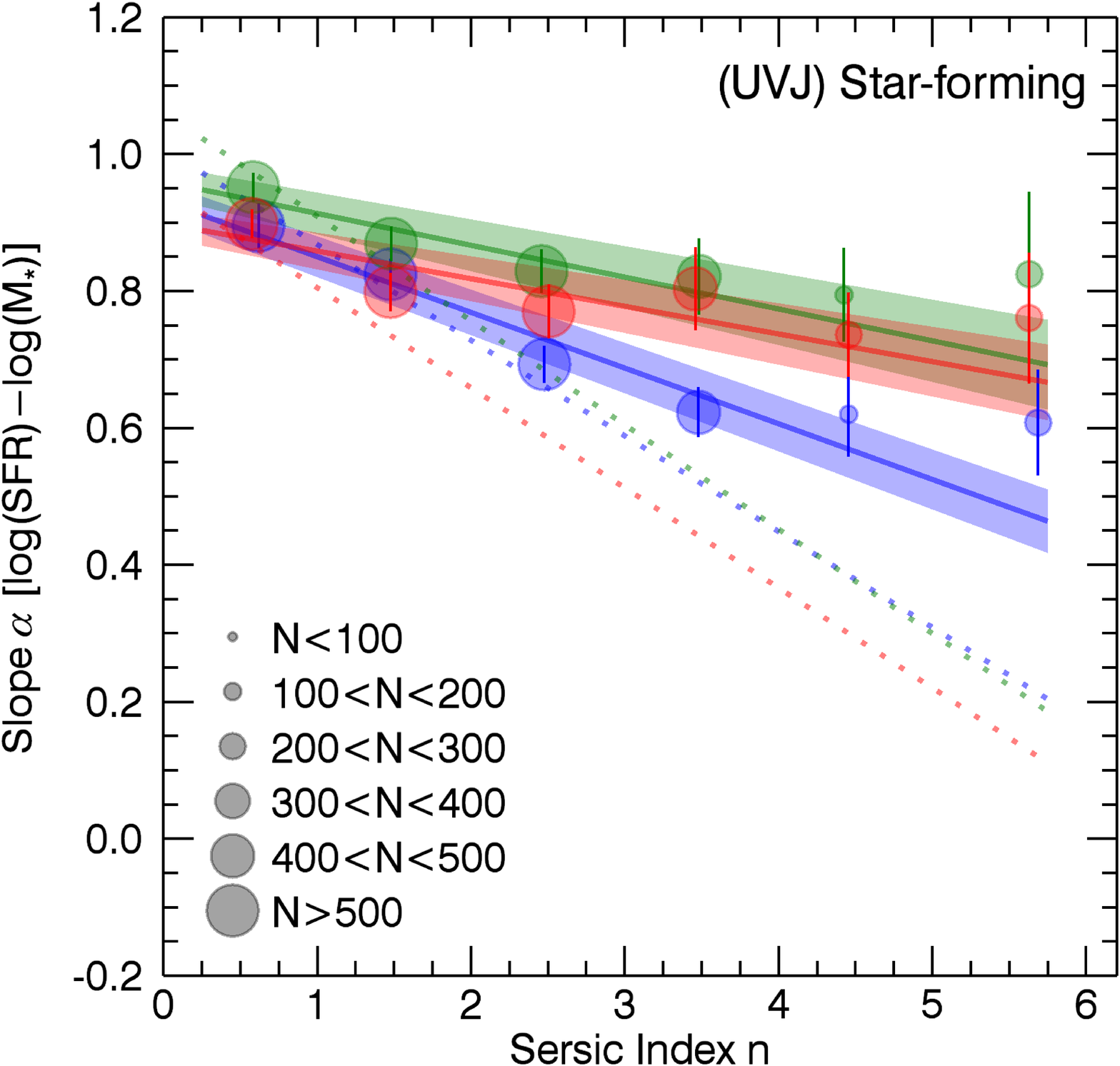}}\hfill
\caption{The best-fit linear slope $\alpha$ of the $\log\mathrm{SFR}-\log\mathrm{M_{\star}}$ relation
above the star-forming mass limits depends strongly on $n$ when both
star-forming and quiescent galaxies are considered.  This trend is weaker
for star-forming galaxies only, and may signify active bulge-formation amongst the most massive galaxies
at $z$$\sim$2, where the stars in the central concentration are younger.
The symbol size represents the number of galaxies $N$.  Dotted lines are the best-fit relation from the
opposite panel for reference.}
\label{fig:slope}
\end{figure*}

Figure~\ref{fig:sersic_sfr} shows the $\log\mathrm{SFR}$--$\log\mathrm{M_{\star}}$ projection 
of Figure~\ref{fig:sersic},
color-coded by $n$.  At all redshifts, we see a clear correlation between SFR/M$_{\star}$ and $n$
when considering both quiescent and star-forming galaxies (left panels).  
Although this result is primarily driven
by the bulge-dominated, low sSFR quiescent population, we do see a weaker trend
for star-forming galaxies only.  The more concentrated the light profile of a galaxy 
(high-$n$), the more likely that galaxy is to have a depressed sSFR relative to the star formation
sequence. 

Assuming $n$ is a proxy for bulge-to-disk ratio,
we see evidence to support the idea of \citet{Abramson14}: the slope of the star formation sequence is correlated
with the build-up of bulges.
We note, however, that even galaxies with $n$=1 exhibit some 
curvature of the star formation sequence, suggesting an additional mechanism acting to suppress star formation in massive
galaxies.

Figure~\ref{fig:slope} shows the slope of $\log\mathrm{SFR}-\log\mathrm{M}_{\star}$ relation measured
above the star-forming mass limits.  The error bars include the uncertainty in the slope measurement itself and $n$,
where the latter is calculated from 100 bootstrap simulations perturbing $n$ by the random errors and repeating
the analysis.
A summary of the best-fit parameters is found in Table~\ref{tab:linfit}.
We find similar general trends when selecting star-forming galaxies by $\log\mathrm{(sSFR)}$$>$-10 yr$^{-1}$.
\citet{Whitaker12b} demonstrated that a steeper slope of the star formation sequence
is measured when selecting blue star-forming galaxies \citep[e.g.,][]{Peng10}.  
Here, we show explicitly that by selecting galaxies
with prominent disks (which often have blue rest-frame optical colors), we measure a 
slope close to unity across the full range in stellar mass and redshift.  We confirm similar
results by \citet{Wuyts11b}, and emphasize the critical importance of understanding selection biases.

\section{Discussion}
\label{sec:discussion}

\begin{figure*}
\includegraphics[width=0.98\linewidth]{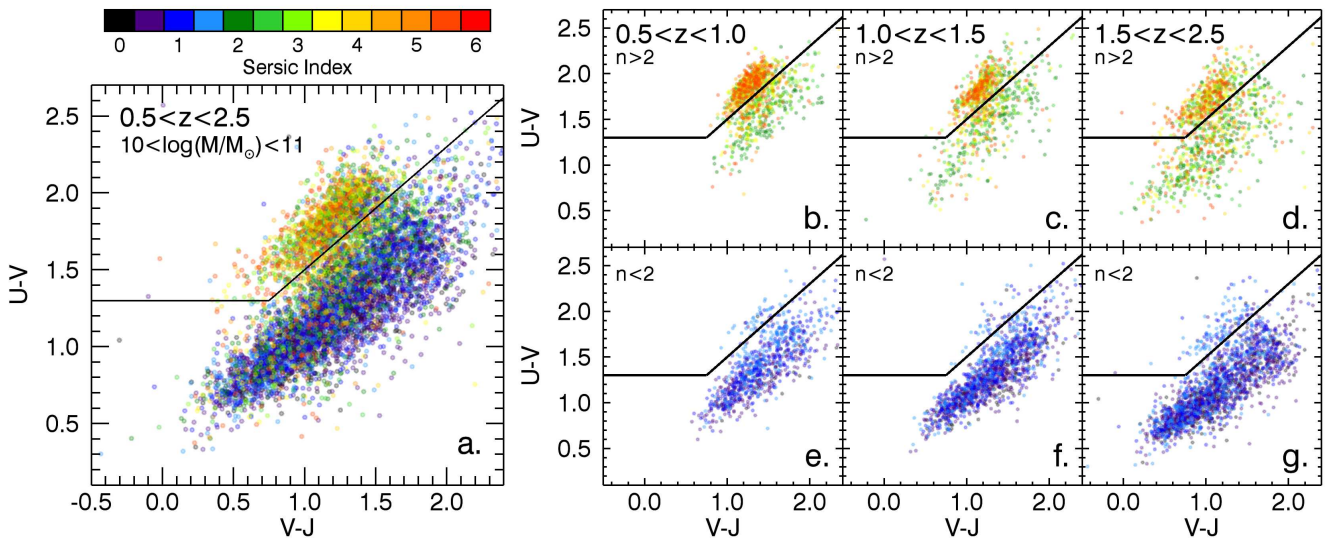}
\caption{The rest-frame $U$--$V$ and  $V$--$J$ colors also reflect a bimodality in the Sersic index (Panel A), with bulge-dominated
galaxies predominatly inhabiting the quiescent region and disk-dominated galaxies in the star-forming region.
There do not exist strong trends within the quiescent and star-forming populations.
The notable exception is the cross-over region between these two populations, where galaxies tend to be bulge-dominated, matching the quiescent 
population, while still residing in the star-forming color-color parameter space at $z$$<$1 (Panel B).
Conversely, at $z$$>$1.5, star-forming galaxies with high S\'{e}rsic indices cover the full color-color parameter space (Panel D).}
\label{fig:uvj}
\end{figure*}

The aim of this letter is to connect rest-frame optical measurements of the
$n$-M$_{\star}$ relation with total sSFRs for a single SFR indicator from a purely empirical standpoint.
We can thereby connect galaxy structure and star formation to understand the observed bimodal distribution of galaxies across cosmic time.
We measure a systematic decrease in the global sSFRs of galaxies with increasing $n$.
  This flattening of the star formation sequence
directly reflects the bimodality of the two stellar populations: quiescent galaxies tend to be
bulge-dominated with significantly lower sSFRs, whereas star-forming galaxies are typically
disk-dominated with sSFRs consistent with the main ridge of the star formation sequence \citep[confirming][]{Wuyts11b}.
The measured SFR-M$_{\star}$ relation for galaxies with $n$=2 tracks the best-fit relations of \citet{Whitaker14b} in Figure~\ref{fig:sersic_sfr}.
Galaxies with $n$$<$2 have a slightly steeper (but still curved) relation, whereas galaxies with $n$$>$2 show a significantly flattened relation.
We also see that $n$ correlates with the color bimodality of the populations (Figure~\ref{fig:uvj}, Panel A).

When considering star-forming galaxies only, pure exponential-disk galaxies
exhibit a higher slope of the $\log\mathrm{SFR}-\log\mathrm{M}_{\star}$ relation 
than those with a significant bulge component.  
This result is somewhat dependent on how star-forming is defined: 
UVJ-selected star-forming galaxies at $z$$<$1 with $n$$>$2 have rest-frame colors 
close to the quiescent region (Panel B, Figure~\ref{fig:uvj}). 
If we shift the quiescent box redward in $V$--$J$ by +0.1--0.3 mag and repeat the 
analysis, we find the same correlation between the slope of the star-formation 
sequence and $n$ at $z$$<$1. This is because high-$n$ galaxies drop out of the 
sample altogether but variations in the slope from $n$=0.5 to $n$=3 remain.
The galaxies that lie in this intermediate region of rest-frame color space may be
an extension of the quiescent population
on the verge of shutting down star formation (e.g., \citet{Belli15, Papovich15}, 
Yano et al. in prep).
They have more concentrated light profiles and lower sSFRs than typical 
star-forming galaxies.

For star-forming galaxies at $z$$>$1, the slope of the star-formation sequence instead 
depends weakly on $n$ (right panel, Figure~\ref{fig:slope}).  The correlation is not 
statistically significant above $n$$>$1.
Shifting the UVJ selection
redward increasingly weakens the trend.  Perhaps the larger photometric errors of quiescent galaxies
at high-$z$ scatter some into the intermediate parameter space, which get removed by this test.  The
$n$$>$2 star-forming galaxies well separated from the UVJ line
show a similar range of colors to $n$$<$2 star-forming galaxies (compare Panels D and G,
Figure~\ref{fig:uvj}). They may also have similar sSFRs, although $n$$>$3 star-forming
galaxies are not prevalent at the highest masses.
Disk-dominated star-forming galaxies are neither preferentially dusty or dust-free; their range in colors in Figure~\ref{fig:uvj}
likely reflects the inclination
angle through which we view the disk \citep{Patel12}.
The process of forming a significantly massive bulge at $z$$>$1 appears to be connected 
to a transformation in the rest-frame optical galaxy structure and color before the 
decrease of the global sSFR of galaxies.  
Panel D in Figure~\ref{fig:uvj} demonstrates that
the galaxy morphology is in place at $z$$\sim$2 before the reddening of the rest-frame colors associated with passive evolution.

The observation that the correlation between the slope of the star-formation sequence and $n$ is
weaker amongst star-forming
galaxies at $z$$>$1 may 
tell us something about bulge formation.  Traditional bulges may already be in place at $z$$<$1.  For 
example, \citet{vanDokkum14} find little growth in the centers of massive galaxies below $z$$<$0.8, but at higher
redshifts there is growth at all radii. Furthermore, \citet{Nelson12} demonstrate that (new) star formation
at $z$$\sim$1 occurs at all radii in exponential disks.  In other words, bulges at $z$$\lesssim$1 may already be composed
of older stars.

The picture is less clear at higher redshifts; it may be that massive galaxies are actively 
building up their bulges.  From Figures~\ref{fig:sersic}-\ref{fig:slope}, 
we see that bulge-dominated galaxies at high-$z$ exist,
but don't seem to have a significantly lower global sSFR.  
If the bulges of massive $z$$\sim$2 galaxies are actively forming, the stars will still
be young and there should not yet be a strong depression in global sSFRs.  
\citet{Choi14} derived stellar ages for a large sample of quiescent galaxies at 0.1$<$$z$$<$0.7
that are consistent with an equivalent single-burst star formation epoch of $z\sim$1.5,
supporting the idea that $z$$\sim$2 is an important epoch for the bulge formation for massive galaxies.
This doesn't however tell us anything about how bulges
grow.  We cannot tell if clumps of stars form insitu and migrate to the 
bulge due to disk instabilities \citep[e.g.,][]{Noguchi99, Elmegreen08}, if they are the 
result of a rich merger history \citep[e.g.,][]{DeLucia11}, or some other mechanism.    

The best way to decompose the star formation histories of $z$$\sim$2 galaxies will be with
future spatially-resolved studies.  For example, Nelson et al. (in prep) find that the
radial H$\alpha$ profiles of galaxies above the ridge of the star-formation sequence are enhanced at all radii, whereas galaxies below have suppressed star formation at all radii. Their results
suggests that a global process regulates gas accretion, and the scatter in the star formation 
sequence is real.
It is however unclear why the amount of gas accretion depends on disk mass instead of total mass.
In this work, we demonstrate that this scatter may be related in part to $n$.
In Figure~\ref{fig:sersic_sfr}, the scatter of the SFRs between 10$<$$\log\mathrm{(M/M_{\odot})}$$<$11 for all galaxies 
ranges from 0.34$\pm$0.09 dex at 1.5$<$$z$$<$2.5, to 0.43$\pm$0.05 dex at 1.0$<$$z$$<$1.5, and 0.48$\pm$0.02 dex at 0.5$<$$z$$<$1.0.
When considering UVJ-selected star-forming galaxies only, these values are 0.14$\pm$0.02 dex at 1.5$<$$z$$<$2.5, 
0.21$\pm$0.05 dex at 1.0$<$$z$$<$1.5, and  0.30$\pm$0.04 dex at 0.5$<$$z$$<$1.0.
Compared to the roughly constant 0.34 dex observed scatter of the
star-formation sequence \citep{Whitaker12b}, we show that this scatter is driven in part by real
differences in $n$.
Random errors will introduce $\sim$0.05 dex 
scatter, and evolution of the average SFR within the redshift bins $\sim$0.11--0.14 dex \citep{Whitaker12b}. 
It may be that a significant portion of the remaining scatter is correlated with variations in $n$ amongst galaxies.
However, we are unable to decompose the relative contributions of intrinsic and systematic scatter at present, and note that the 24$\mu$m-L$_{\mathrm{IR}}$ 
conversion itself could introduce as much 0.25 dex of scatter \citep{Wuyts11a}.

\begin{table*}[t]
\centering
\begin{threeparttable}
  \caption{Linear fits to the slope $\alpha$ of the $\log\mathrm{SFR}-\log\mathrm{M}_{\star}$
    relation vs. S\'{e}rsic index}\label{tab:linfit}
    \begin{tabular}{lcccc}
	  \hline \hline
      & \multicolumn{2}{c}{All Galaxies} & \multicolumn{2}{c}{(UVJ) Star-forming} \\
      \cline{2-3} \cline{4-5}
      redshift range ~~~~~~ &  ~~~~~~~~~$a$~~~~~~~~~ & ~~~~~~~~~$b$~~~~~~~~~ & ~~~~~~~~~$a$~~~~~~~~~ & ~~~~~~~~~$b$~~~~~~~~~ \\
      \hline
      \noalign{\smallskip}
      $0.5<z<1.0$  & $1.01\pm0.03$ & $-0.14\pm0.01$  & $0.93\pm0.03$ & $-0.08\pm0.01$ \\
      $1.0<z<1.5$  & $1.06\pm0.03$ & $-0.15\pm0.01$  & $0.96\pm0.03$ & $-0.05\pm0.02$ \\
      $2.5<z<2.5$  & $0.95\pm0.03$ & $-0.15\pm0.01$  & $0.90\pm0.03$ & $-0.04\pm0.01$ \\
      \noalign{\smallskip}
      \hline
      \noalign{\smallskip}
    \end{tabular}
    \begin{tablenotes}
      \small
    \item \emph{Notes.} Linear fit coefficients parameterizing the evolution of the slope, $\alpha$, of the 
    	$\log\mathrm{SFR}-\log\mathrm{M_{\star}}$ relation as a function of S\'{e}rsic index, n, where $\alpha$=a+b$n$.
    \end{tablenotes}
  \end{threeparttable}
\end{table*}

%%=== Fig 4

These results support the idea that we are witnessing the rapid build-up of bulges at $z\sim$2 in massive galaxies.
A prominent bulge may be an important condition for quenching to occur \citep[e.g.,][]{Bell12}, at least
in massive galaxies. \citet{Franx08} demonstrated that surface density and inferred velocity dispersion are 
better correlated with sSFR and color than stellar mass.  We will explore these dependencies
in an upcoming paper (Whitaker et al. in preparation).

\begin{acknowledgements}
We thank the anonymous referee for a highly constructive report.
%The authors are grateful to the many colleagues who have
%provided public data and catalogs in the five deep 3D-HST fields; high redshift galaxy science
%has thrived owing to this gracious mindset and the TACs and the Observatory Directors who have encouraged this.
This work is based on observations taken by the 3D-HST Treasury Program (GO 12177 and 12328)
with the NASA/ESA HST, which is operated by the Associations of Universities for Resarch in Astronomy, Inc.,
under NASA contract NAS5-26555.
This research was supported by an appointment to the NASA Postdoctoral Program at 
the Goddard Space Flight Center, administered by Oak 
Ridge Associated Universities through a contract with NASA.
\end{acknowledgements}

\clearpage

\addcontentsline{toc}{chapter}{\numberline {}{\sc References}}


\begin{thebibliography}{45}
\expandafter\ifx\csname natexlab\endcsname\relax\def\natexlab#1{#1}\fi

\bibitem[{{Abramson} {et~al.}(2014){Abramson}, {Kelson}, {Dressler},
  {Poggianti}, {Gladders}, {Oemler}, \& {Vulcani}}]{Abramson14}
{Abramson}, L.~E., {Kelson}, D.~D., {Dressler}, A., {et~al.} 2014, \apjl, 785,
  L36

\bibitem[{{Barro} {et~al.}(2013){Barro}, {Faber}, {P{\'e}rez-Gonz{\'a}lez},
  {Koo}, {Williams}, {Kocevski}, {Trump}, {Mozena}, {McGrath}, {van der Wel},
  {Wuyts}, {Bell}, {Croton}, {Ceverino}, {Dekel}, {Ashby}, {Cheung},
  {Ferguson}, {Fontana}, {Fang}, {Giavalisco}, {Grogin}, {Guo}, {Hathi},
  {Hopkins}, {Huang}, {Koekemoer}, {Kartaltepe}, {Lee}, {Newman}, {Porter},
  {Primack}, {Ryan}, {Rosario}, {Somerville}, {Salvato}, \& {Hsu}}]{Barro13}
{Barro}, G., {Faber}, S.~M., {P{\'e}rez-Gonz{\'a}lez}, P.~G., {et~al.} 2013,
  \apj, 765, 104

\bibitem[{{Bell} {et~al.}(2012){Bell}, {van der Wel}, {Papovich}, {Kocevski},
  {Lotz}, {McIntosh}, {Kartaltepe}, {Faber}, {Ferguson}, {Koekemoer}, {Grogin},
  {Wuyts}, {Cheung}, {Conselice}, {Dekel}, {Dunlop}, {Giavalisco},
  {Herrington}, {Koo}, {McGrath}, {de Mello}, {Rix}, {Robaina}, \&
  {Williams}}]{Bell12}
{Bell}, E.~F., {van der Wel}, A., {Papovich}, C., {et~al.} 2012, \apj, 753, 167

\bibitem[{{Belli} {et~al.}(2015){Belli}, {Newman}, \& {Ellis}}]{Belli15}
{Belli}, S., {Newman}, A.~B., \& {Ellis}, R.~S. 2015, \apj, 799, 206

\bibitem[{{Brammer} {et~al.}(2012){Brammer}, {van Dokkum}, {Franx},
  {Fumagalli}, {Patel}, {Rix}, {Skelton}, {Kriek}, {Nelson}, {Schmidt},
  {Bezanson}, {da Cunha}, {Erb}, {Fan}, {F{\"o}rster Schreiber}, {Illingworth},
  {Labb{\'e}}, {Leja}, {Lundgren}, {Magee}, {Marchesini}, {McCarthy},
  {Momcheva}, {Muzzin}, {Quadri}, {Steidel}, {Tal}, {Wake}, {Whitaker}, \&
  {Williams}}]{Brammer12}
{Brammer}, G.~B., {van Dokkum}, P.~G., {Franx}, M., {et~al.} 2012, \apjs, 200,
  13

\bibitem[{{Brinchmann} {et~al.}(2004){Brinchmann}, {Charlot}, {White},
  {Tremonti}, {Kauffmann}, {Heckman}, \& {Brinkmann}}]{Brinchmann04}
{Brinchmann}, J., {Charlot}, S., {White}, S.~D.~M., {et~al.} 2004, \mnras, 351,
  1151

\bibitem[{{Bruce} {et~al.}(2014){Bruce}, {Dunlop}, {McLure}, {Cirasuolo},
  {Buitrago}, {Bowler}, {Targett}, {Bell}, {McIntosh}, {Dekel}, {Faber},
  {Ferguson}, {Grogin}, {Hartley}, {Kocevski}, {Koekemoer}, {Koo}, \&
  {McGrath}}]{Bruce14}
{Bruce}, V.~A., {Dunlop}, J.~S., {McLure}, R.~J., {et~al.} 2014, \mnras, 444,
  1660

\bibitem[{{Caon} {et~al.}(1993){Caon}, {Capaccioli}, \& {D'Onofrio}}]{Caon93}
{Caon}, N., {Capaccioli}, M., \& {D'Onofrio}, M. 1993, \mnras, 265, 1013

\bibitem[{{Chabrier}(2003)}]{Chabrier}
{Chabrier}, G. 2003, \pasp, 115, 763

\bibitem[{{Choi} {et~al.}(2014){Choi}, {Conroy}, {Moustakas}, {Graves},
  {Holden}, {Brodwin}, {Brown}, \& {van Dokkum}}]{Choi14}
{Choi}, J., {Conroy}, C., {Moustakas}, J., {et~al.} 2014, \apj, 792, 95

\bibitem[{{De Lucia} {et~al.}(2011){De Lucia}, {Fontanot}, {Wilman}, \&
  {Monaco}}]{DeLucia11}
{De Lucia}, G., {Fontanot}, F., {Wilman}, D., {et~al.} 2011, \mnras, 414, 1439

\bibitem[{{de Vaucouleurs}(1948)}]{deVaucouleur}
{de Vaucouleurs}, G. 1948, Annales d'Astrophysique, 11, 247

\bibitem[{{Dickinson} \& {FIDEL Team}(2007)}]{Dickinson07}
{Dickinson}, M., \& {FIDEL Team}. 2007, in Bulletin of the American
  Astronomical Society, Vol.~39, American Astronomical Society Meeting
  Abstracts, 822

\bibitem[{{Dickinson} {et~al.}(2003){Dickinson}, {Papovich}, {Ferguson}, \&
  {Budav{\'a}ri}}]{Dickinson03}
{Dickinson}, M., {Papovich}, C., {Ferguson}, H.~C., {et~al.} 2003, \apj, 587,
  25

\bibitem[{{Donley} {et~al.}(2012){Donley}, {Koekemoer}, {Brusa}, {Capak},
  {Cardamone}, {Civano}, {Ilbert}, {Impey}, {Kartaltepe}, {Miyaji}, {Salvato},
  {Sanders}, {Trump}, \& {Zamorani}}]{Donley12}
{Donley}, J.~L., {Koekemoer}, A.~M., {Brusa}, M., {et~al.} 2012, \apj, 748, 142

\bibitem[{{Elmegreen} {et~al.}(2008){Elmegreen}, {Bournaud}, \&
  {Elmegreen}}]{Elmegreen08}
{Elmegreen}, B.~G., {Bournaud}, F., \& {Elmegreen}, D.~M. 2008, \apj, 688, 67

\bibitem[{{Franx} {et~al.}(2008){Franx}, {van Dokkum}, {Schreiber}, {Wuyts},
  {Labb{\'e}}, \& {Toft}}]{Franx08}
{Franx}, M., {van Dokkum}, P.~G., {Schreiber}, N.~M.~F., {et~al.} 2008, \apj,
  688, 770

\bibitem[{{Grogin} {et~al.}(2011){Grogin}, {Kocevski}, {Faber}, {Ferguson},
  {Koekemoer}, {Riess}, {Acquaviva}, {Alexander}, {Almaini}, {Ashby}, {Barden},
  {Bell}, {Bournaud}, {Brown}, {Caputi}, {Casertano}, {Cassata}, {Castellano},
  {Challis}, {Chary}, {Cheung}, {Cirasuolo}, {Conselice}, {Roshan Cooray},
  {Croton}, {Daddi}, {Dahlen}, {Dav{\'e}}, {de Mello}, {Dekel}, {Dickinson},
  {Dolch}, {Donley}, {Dunlop}, {Dutton}, {Elbaz}, {Fazio}, {Filippenko},
  {Finkelstein}, {Fontana}, {Gardner}, {Garnavich}, {Gawiser}, {Giavalisco},
  {Grazian}, {Guo}, {Hathi}, {H{\"a}ussler}, {Hopkins}, {Huang}, {Huang},
  {Jha}, {Kartaltepe}, {Kirshner}, {Koo}, {Lai}, {Lee}, {Li}, {Lotz}, {Lucas},
  {Madau}, {McCarthy}, {McGrath}, {McIntosh}, {McLure}, {Mobasher},
  {Moustakas}, {Mozena}, {Nandra}, {Newman}, {Niemi}, {Noeske}, {Papovich},
  {Pentericci}, {Pope}, {Primack}, {Rajan}, {Ravindranath}, {Reddy}, {Renzini},
  {Rix}, {Robaina}, {Rodney}, {Rosario}, {Rosati}, {Salimbeni}, {Scarlata},
  {Siana}, {Simard}, {Smidt}, {Somerville}, {Spinrad}, {Straughn}, {Strolger},
  {Telford}, {Teplitz}, {Trump}, {van der Wel}, {Villforth}, {Wechsler},
  {Weiner}, {Wiklind}, {Wild}, {Wilson}, {Wuyts}, {Yan}, \& {Yun}}]{Grogin11}
{Grogin}, N.~A., {Kocevski}, D.~D., {Faber}, S.~M., {et~al.} 2011, \apjs, 197,
  35

\bibitem[{{Hayward} {et~al.}(2014){Hayward}, {Lanz}, {Ashby}, {Fazio},
  {Hernquist}, {Mart{\'{\i}}nez-Galarza}, {Noeske}, {Smith}, {Wuyts}, \&
  {Zezas}}]{Hayward14}
{Hayward}, C.~C., {Lanz}, L., {Ashby}, M.~L.~N., {et~al.} 2014, \mnras, 445,
  1598

\bibitem[{{Kennicutt}(1998)}]{Kennicutt98}
{Kennicutt}, Jr., R.~C. 1998, \araa, 36, 189

\bibitem[{{Koekemoer} {et~al.}(2011){Koekemoer}, {Faber}, {Ferguson}, {Grogin},
  {Kocevski}, {Koo}, {Lai}, {Lotz}, {Lucas}, {McGrath}, {Ogaz}, {Rajan},
  {Riess}, {Rodney}, {Strolger}, {Casertano}, {Castellano}, {Dahlen},
  {Dickinson}, {Dolch}, {Fontana}, {Giavalisco}, {Grazian}, {Guo}, {Hathi},
  {Huang}, {van der Wel}, {Yan}, {Acquaviva}, {Alexander}, {Almaini}, {Ashby},
  {Barden}, {Bell}, {Bournaud}, {Brown}, {Caputi}, {Cassata}, {Challis},
  {Chary}, {Cheung}, {Cirasuolo}, {Conselice}, {Roshan Cooray}, {Croton},
  {Daddi}, {Dav{\'e}}, {de Mello}, {de Ravel}, {Dekel}, {Donley}, {Dunlop},
  {Dutton}, {Elbaz}, {Fazio}, {Filippenko}, {Finkelstein}, {Frazer}, {Gardner},
  {Garnavich}, {Gawiser}, {Gruetzbauch}, {Hartley}, {H{\"a}ussler},
  {Herrington}, {Hopkins}, {Huang}, {Jha}, {Johnson}, {Kartaltepe},
  {Khostovan}, {Kirshner}, {Lani}, {Lee}, {Li}, {Madau}, {McCarthy},
  {McIntosh}, {McLure}, {McPartland}, {Mobasher}, {Moreira}, {Mortlock},
  {Moustakas}, {Mozena}, {Nandra}, {Newman}, {Nielsen}, {Niemi}, {Noeske},
  {Papovich}, {Pentericci}, {Pope}, {Primack}, {Ravindranath}, {Reddy},
  {Renzini}, {Rix}, {Robaina}, {Rosario}, {Rosati}, {Salimbeni}, {Scarlata},
  {Siana}, {Simard}, {Smidt}, {Snyder}, {Somerville}, {Spinrad}, {Straughn},
  {Telford}, {Teplitz}, {Trump}, {Vargas}, {Villforth}, {Wagner}, {Wandro},
  {Wechsler}, {Weiner}, {Wiklind}, {Wild}, {Wilson}, {Wuyts}, \&
  {Yun}}]{Koekemoer11}
{Koekemoer}, A.~M., {Faber}, S.~M., {Ferguson}, H.~C., {et~al.} 2011, \apjs,
  197, 36

\bibitem[{{Kriek} {et~al.}(2009){Kriek}, {van Dokkum}, {Franx}, {Illingworth},
  \& {Magee}}]{Kriek09b}
{Kriek}, M., {van Dokkum}, P.~G., {Franx}, M., {et~al.} 2009, \apjl, 705, L71

\bibitem[{{Labb{\'e}} {et~al.}(2010){Labb{\'e}}, {Gonz{\'a}lez}, {Bouwens},
  {Illingworth}, {Oesch}, {van Dokkum}, {Carollo}, {Franx}, {Stiavelli},
  {Trenti}, {Magee}, \& {Kriek}}]{Labbe10}
{Labb{\'e}}, I., {Gonz{\'a}lez}, V., {Bouwens}, R.~J., {et~al.} 2010, \apjl,
  708, L26

\bibitem[{{Lang} {et~al.}(2014){Lang}, {Wuyts}, {Somerville}, {F{\"o}rster
  Schreiber}, {Genzel}, {Bell}, {Brammer}, {Dekel}, {Faber}, {Ferguson},
  {Grogin}, {Kocevski}, {Koekemoer}, {Lutz}, {McGrath}, {Momcheva}, {Nelson},
  {Primack}, {Rosario}, {Skelton}, {Tacconi}, {van Dokkum}, \&
  {Whitaker}}]{Lang14}
{Lang}, P., {Wuyts}, S., {Somerville}, R.~S., {et~al.} 2014, \apj, 788, 11

\bibitem[{{Nelson} {et~al.}(2012){Nelson}, {van Dokkum}, {Brammer},
  {F{\"o}rster Schreiber}, {Franx}, {Fumagalli}, {Patel}, {Rix}, {Skelton},
  {Bezanson}, {Da Cunha}, {Kriek}, {Labbe}, {Lundgren}, {Quadri}, \&
  {Schmidt}}]{Nelson12}
{Nelson}, E.~J., {van Dokkum}, P.~G., {Brammer}, G., {et~al.} 2012, \apjl, 747,
  L28

\bibitem[{{Noeske} {et~al.}(2007){Noeske}, {Weiner}, {Faber}, {Papovich},
  {Koo}, {Somerville}, {Bundy}, {Conselice}, {Newman}, {Schiminovich}, {Le
  Floc'h}, {Coil}, {Rieke}, {Lotz}, {Primack}, {Barmby}, {Cooper}, {Davis},
  {Ellis}, {Fazio}, {Guhathakurta}, {Huang}, {Kassin}, {Martin}, {Phillips},
  {Rich}, {Small}, {Willmer}, \& {Wilson}}]{Noeske07a}
{Noeske}, K.~G., {Weiner}, B.~J., {Faber}, S.~M., {et~al.} 2007, \apjl, 660,
  L43

\bibitem[{{Noguchi}(1999)}]{Noguchi99}
{Noguchi}, M. 1999, \apj, 514, 77

\bibitem[{{Papovich} {et~al.}(2015){Papovich}, {Labb{\'e}}, {Quadri}, {Tilvi},
  {Behroozi}, {Bell}, {Glazebrook}, {Spitler}, {Straatman}, {Tran}, {Cowley},
  {Dav{\'e}}, {Dekel}, {Dickinson}, {Ferguson}, {Finkelstein}, {Gawiser},
  {Inami}, {Faber}, {Kacprzak}, {Kawinwanichakij}, {Kocevski}, {Koekemoer},
  {Koo}, {Kurczynski}, {Lotz}, {Lu}, {Lucas}, {McIntosh}, {Mehrtens},
  {Mobasher}, {Monson}, {Morrison}, {Nanayakkara}, {Persson}, {Salmon},
  {Simons}, {Tomczak}, {van Dokkum}, {Weiner}, \& {Willner}}]{Papovich15}
{Papovich}, C., {Labb{\'e}}, I., {Quadri}, R., {et~al.} 2015, \apj, 803, 26

\bibitem[{{Patel} {et~al.}(2012){Patel}, {Holden}, {Kelson}, {Franx}, {van der
  Wel}, \& {Illingworth}}]{Patel12}
{Patel}, S.~G., {Holden}, B.~P., {Kelson}, D.~D., {et~al.} 2012, \apjl, 748,
  L27

\bibitem[{{Peng} {et~al.}(2010){Peng}, {Lilly}, {Kova{\v c}}, {Bolzonella},
  {Pozzetti}, {Renzini}, {Zamorani}, {Ilbert}, {Knobel}, {Iovino}, {Maier},
  {Cucciati}, {Tasca}, {Carollo}, {Silverman}, {Kampczyk}, {de Ravel},
  {Sanders}, {Scoville}, {Contini}, {Mainieri}, {Scodeggio}, {Kneib}, {Le
  F{\`e}vre}, {Bardelli}, {Bongiorno}, {Caputi}, {Coppa}, {de la Torre},
  {Franzetti}, {Garilli}, {Lamareille}, {Le Borgne}, {Le Brun}, {Mignoli},
  {Perez Montero}, {Pello}, {Ricciardelli}, {Tanaka}, {Tresse}, {Vergani},
  {Welikala}, {Zucca}, {Oesch}, {Abbas}, {Barnes}, {Bordoloi}, {Bottini},
  {Cappi}, {Cassata}, {Cimatti}, {Fumana}, {Hasinger}, {Koekemoer},
  {Leauthaud}, {Maccagni}, {Marinoni}, {McCracken}, {Memeo}, {Meneux}, {Nair},
  {Porciani}, {Presotto}, \& {Scaramella}}]{Peng10}
{Peng}, Y.-j., {Lilly}, S.~J., {Kova{\v c}}, K., {et~al.} 2010, \apj, 721, 193

\bibitem[{{Sanders} {et~al.}(2007){Sanders}, {Salvato}, {Aussel}, {Ilbert},
  {Scoville}, {Surace}, {Frayer}, {Sheth}, {Helou}, {Brooke}, {Bhattacharya},
  {Yan}, {Kartaltepe}, {Barnes}, {Blain}, {Calzetti}, {Capak}, {Carilli},
  {Carollo}, {Comastri}, {Daddi}, {Ellis}, {Elvis}, {Fall}, {Franceschini},
  {Giavalisco}, {Hasinger}, {Impey}, {Koekemoer}, {Le F{\`e}vre}, {Lilly},
  {Liu}, {McCracken}, {Mobasher}, {Renzini}, {Rich}, {Schinnerer}, {Shopbell},
  {Taniguchi}, {Thompson}, {Urry}, \& {Williams}}]{Sanders07}
{Sanders}, D.~B., {Salvato}, M., {Aussel}, H., {et~al.} 2007, \apjs, 172, 86

\bibitem[{{Schreiber} {et~al.}(2015){Schreiber}, {Pannella}, {Elbaz},
  {B{\'e}thermin}, {Inami}, {Dickinson}, {Magnelli}, {Wang}, {Aussel}, {Daddi},
  {Juneau}, {Shu}, {Sargent}, {Buat}, {Faber}, {Ferguson}, {Giavalisco},
  {Koekemoer}, {Magdis}, {Morrison}, {Papovich}, {Santini}, \&
  {Scott}}]{Schreiber15}
{Schreiber}, C., {Pannella}, M., {Elbaz}, D., {et~al.} 2015, \aap, 575, A74

\bibitem[{{Sersic}(1968)}]{Sersic68}
{Sersic}, J.~L. 1968, {Atlas de galaxias australes}, ed. {Sersic, J.~L.}

\bibitem[{{Skelton} {et~al.}(2014){Skelton}, {Whitaker}, {Momcheva}, {Brammer},
  {van Dokkum}, {Labbe}, {Franx}, {van der Wel}, {Bezanson}, {Da Cunha},
  {Fumagalli}, {Foerster Schreiber}, {Kriek}, {Leja}, {Lundgren}, {Magee},
  {Marchesini}, {Maseda}, {Nelson}, {Oesch}, {Pacifici}, {Patel}, {Price},
  {Rix}, {Tal}, {Wake}, \& {Wuyts}}]{Skelton14}
{Skelton}, R.~E., {Whitaker}, K.~E., {Momcheva}, I.~G., {et~al.} 2014, ArXiv
  e-prints

\bibitem[{{Tal} {et~al.}(2014){Tal}, {Dekel}, {Oesch}, {Muzzin}, {Brammer},
  {van Dokkum}, {Franx}, {Illingworth}, {Leja}, {Magee}, {Marchesini},
  {Momcheva}, {Nelson}, {Patel}, {Quadri}, {Rix}, {Skelton}, {Wake}, \&
  {Whitaker}}]{Tal14}
{Tal}, T., {Dekel}, A., {Oesch}, P., {et~al.} 2014, ArXiv e-prints

\bibitem[{{Utomo} {et~al.}(2014){Utomo}, {Kriek}, {Labb{\'e}}, {Conroy}, \&
  {Fumagalli}}]{Utomo14}
{Utomo}, D., {Kriek}, M., {Labb{\'e}}, I., {et~al.} 2014, \apjl, 783, L30

\bibitem[{{van der Wel} {et~al.}(2012){van der Wel}, {Bell}, {H{\"a}ussler},
  {McGrath}, {Chang}, {Guo}, {McIntosh}, {Rix}, {Barden}, {Cheung}, {Faber},
  {Ferguson}, {Galametz}, {Grogin}, {Hartley}, {Kartaltepe}, {Kocevski},
  {Koekemoer}, {Lotz}, {Mozena}, {Peth}, \& {Peng}}]{vanderWel12}
{van der Wel}, A., {Bell}, E.~F., {H{\"a}ussler}, B., {et~al.} 2012, \apjs,
  203, 24

\bibitem[{{van der Wel} {et~al.}(2014){van der Wel}, {Franx}, {van Dokkum},
  {Skelton}, {Momcheva}, {Whitaker}, {Brammer}, {Bell}, {Rix}, {Wuyts},
  {Ferguson}, {Holden}, {Barro}, {Koekemoer}, {Chang}, {McGrath},
  {H{\"a}ussler}, {Dekel}, {Behroozi}, {Fumagalli}, {Leja}, {Lundgren},
  {Maseda}, {Nelson}, {Wake}, {Patel}, {Labb{\'e}}, {Faber}, {Grogin}, \&
  {Kocevski}}]{vanderWel14}
{van der Wel}, A., {Franx}, M., {van Dokkum}, P.~G., {et~al.} 2014, \apj, 788,
  28

\bibitem[{{van Dokkum} {et~al.}(2014){van Dokkum}, {Bezanson}, {van der Wel},
  {Nelson}, {Momcheva}, {Skelton}, {Whitaker}, {Brammer}, {Conroy},
  {F{\"o}rster Schreiber}, {Fumagalli}, {Kriek}, {Labb{\'e}}, {Leja},
  {Marchesini}, {Muzzin}, {Oesch}, \& {Wuyts}}]{vanDokkum14}
{van Dokkum}, P.~G., {Bezanson}, R., {van der Wel}, A., {et~al.} 2014, \apj,
  791, 45

\bibitem[{{Whitaker} {et~al.}(2014){Whitaker}, {Franx}, {Leja}, {van Dokkum},
  {Henry}, {Skelton}, {Fumagalli}, {Momcheva}, {Brammer}, {Labbe}, {Nelson}, \&
  {Rigby}}]{Whitaker14b}
{Whitaker}, K.~E., {Franx}, M., {Leja}, J., {et~al.} 2014, ArXiv e-prints

\bibitem[{{Whitaker} {et~al.}(2012{\natexlab{a}}){Whitaker}, {Kriek}, {van
  Dokkum}, {Bezanson}, {Brammer}, {Franx}, \& {Labb{\'e}}}]{Whitaker12a}
{Whitaker}, K.~E., {Kriek}, M., {van Dokkum}, P.~G., {et~al.}
  2012{\natexlab{a}}, \apj, 745, 179

\bibitem[{{Whitaker} {et~al.}(2012{\natexlab{b}}){Whitaker}, {van Dokkum},
  {Brammer}, \& {Franx}}]{Whitaker12b}
{Whitaker}, K.~E., {van Dokkum}, P.~G., {Brammer}, G., {et~al.}
  2012{\natexlab{b}}, \apjl, 754, L29

\bibitem[{{Williams} {et~al.}(2010){Williams}, {Quadri}, {Franx}, {van Dokkum},
  {Toft}, {Kriek}, \& {Labb{\'e}}}]{Williams10}
{Williams}, R.~J., {Quadri}, R.~F., {Franx}, M., {et~al.} 2010, \apj, 713, 738

\bibitem[{{Wuyts} {et~al.}(2011{\natexlab{a}}){Wuyts}, {F{\"o}rster Schreiber},
  {Lutz}, {Nordon}, {Berta}, {Altieri}, {Andreani}, {Aussel}, {Bongiovanni},
  {Cepa}, {Cimatti}, {Daddi}, {Elbaz}, {Genzel}, {Koekemoer}, {Magnelli},
  {Maiolino}, {McGrath}, {P{\'e}rez Garc{\'{\i}}a}, {Poglitsch}, {Popesso},
  {Pozzi}, {Sanchez-Portal}, {Sturm}, {Tacconi}, \& {Valtchanov}}]{Wuyts11a}
{Wuyts}, S., {F{\"o}rster Schreiber}, N.~M., {Lutz}, D., {et~al.}
  2011{\natexlab{a}}, \apj, 738, 106

\bibitem[{{Wuyts} {et~al.}(2011{\natexlab{b}}){Wuyts}, {F{\"o}rster Schreiber},
  {van der Wel}, {Magnelli}, {Guo}, {Genzel}, {Lutz}, {Aussel}, {Barro},
  {Berta}, {Cava}, {Graci{\'a}-Carpio}, {Hathi}, {Huang}, {Kocevski},
  {Koekemoer}, {Lee}, {Le Floc'h}, {McGrath}, {Nordon}, {Popesso}, {Pozzi},
  {Riguccini}, {Rodighiero}, {Saintonge}, \& {Tacconi}}]{Wuyts11b}
{Wuyts}, S., {F{\"o}rster Schreiber}, N.~M., {van der Wel}, A., {et~al.}
  2011{\natexlab{b}}, \apj, 742, 96

\end{thebibliography}
\end{document}